\newcommand{\PLA}[3]{Phys.\ Lett.\ A\ {\bf #1},\ #2 (#3)}

\newcommand{\PRL}[3]{Phys.\ Rev.\ Lett.\ {\bf #1},\ #2 (#3)}

\newcommand{\NAT}[3]{Nature\ {\bf #1},\ #2 (#3)}

\newcommand{\PRA}[3]{Phys.\ Rev.\ A\ {\bf #1},\ #2 (#3)}

\newcommand{\JPA}[3]{J.\ Phys.\ A:\ Math.\ Gen.\ {\bf #1},\ #2 (#3)}

\documentclass[aps,showkeys,showpacs,amsmath,amssymb]{revtex4}
\usepackage{dcolumn}
\usepackage{longtable}
\begin{document}
\title{Perfect Teleportation, Quantum state sharing and Superdense Coding through a Genuinely Entangled Five-qubit State}
\author{Sreraman Muralidharan}
\email{sreraman@loyolacollege.edu} \affiliation{Loyola College,
Nungambakkam, Chennai - 600 034, India}
\author{Prasanta K. Panigrahi}
\email{prasanta@prl.res.in} 
\affiliation{Indian Institute of Science Education and Research (IISER) Kolkata, Salt Lake, Kolkata - 700106, India}
\affiliation{Physical Research
Laboratory, Navrangpura, Ahmedabad - 380 009, India}

\begin{abstract}We investigate the usefulness of  a recently introduced five qubit
state by Brown $\it et \ al. \normalfont$ \cite{Brown} for quantum teleportation, quantum state sharing
and superdense coding. It is shown that this state can
be utilized for perfect teleportation of arbitrary single and two
qubit systems. We devise various schemes for quantum state sharing of an arbitrary 
single and two particle state via cooperative teleportation. 
We later show that this state can be used for
superdense coding as well. It is found that five classical bits
can be sent by sending only three quantum bits.
\end{abstract}
\maketitle

\section{Introduction}
Entanglement is central to all branches of quantum computation and
information. This counter-intuitive feature has been used to
achieve what would be impossible in classical physics.
Characterization and classification of multi-partite entangled
states is not yet firmly established \cite{Plenio}. Quantum
teleportation is an important ingredient in distributed quantum
networks, which exploit entanglement for transferring a quantum
state between two or more parties. It also serves as an elementary
operation in quantum computers and in a number of quantum
communication protocols. It has been achieved experimentally in
different quantum systems \cite{Bou,Riebe,Barrett} and over long
distances, inside \cite{Marc} and outside \cite{Ursin} laboratory
conditions.

Quantum teleportation is a technique for transfer of information
between parties, using a distributed entangled state and a
classical communication channel \cite{Nielsen}. Existence of long
range correlations assist in information transfer by a sender,
unaware of the information to be sent, as well as the destination.
Teleportation of an arbitrary single qubit, $|\psi\rangle_a =
\alpha|0\rangle+\beta|1\rangle$,\ (with $|\alpha|^2+|\beta|^2 =
1$) through an entangled channel of EPR pair between the sender
and receiver was first demonstrated by Bennett {\it et al.}
\cite{Bennett}. Superdense coding is another spectacular
application of quantum information theory, receiving significant
attention in recent times. It shares a close relationship with
quantum teleportation \cite{Werner}. Multi-partite entangled
states, namely the prototype- GHZ states \cite{Karlsson},
generalized W states \cite{Agrawal,Gorbachev} and the cluster
states \cite{Hans,Robert}, have also been exploited for carrying
out teleportation and superdense coding. With the increment in
number of states, the complexity involved increases manifold due
to scarce knowledge regarding characterization of multi-partite
entanglement.

The way in which a given shared multi-particle state is entangled
plays a pivotal role in deciding the suitability of the state for
teleportation. For instance, it is well known that the normal W
states are not useful for perfect teleportation. However,
assigning suitable weights and relative phases to individual terms
of W state makes it suitable for perfect teleportation and
superdense coding \cite{Agrawal}. These modified W states are
unitarily connected to GHZ states \cite{Li}.

Though many type of states have been used for teleporting an
arbitrary one qubit state, very few known states are capable of
teleporting an arbitrary two-qubit state. Even the multi-qubit GHZ
and the generalized W state cannot be used for this purpose. All
the states, that are known to be useful for this purpose are
essentially four qubit states \cite{Rigolin1,Yeo}. Higher even
dimensional generalizations have been explicated, which allow
teleporting N qubit system using N Bell states \cite{Rigolin2}. In
this paper, we describe a five-qubit state that can be used for
perfect teleportation of both a one qubit as well as an arbitrary
two qubit state and discuss its advantages over the previously
known states. It is found suitable to carry out maximal
teleportation and maximal superdense coding, satisfying the
definition of Task-oriented Maximally Entangled States
\cite{Pankaj}. In addition to this, a new five partite cluster state
was introduced in \cite{Pankaj}, for perfect teleportation and superdense coding.
Five qubit entangled states play a key role in quantum information processsing tasks and it is the 
threshold number of qubits required for Quantum error correction \cite{Bennetch, Laflamme}. Quantum mechanical entanglement of five particles was
achieved using the spontaneous parametric down conversion (SPDC) as 
a source of entangled photons \cite{Zhao}. Teleportation was carried out experimentally 
using the five particle entangled GHZ state \cite{Zhao}. Recently, Brown {\it et al.}, \cite{Brown} arrived at a maximally
entangled five-qubit state through an extensive numerical
optimization procedure. This has the form : 

\begin{equation}
|\psi_{5}\rangle=\frac{1}{2}(|001\rangle|\phi_{-}\rangle+|010\rangle|\psi_{-}\rangle
+|100\rangle|\phi_{+}\rangle+|111\rangle|\psi_{+}\rangle),
\end{equation}

where, $|\psi_\pm\rangle=\frac{1}{\sqrt{2}}(|00\rangle\pm|11\rangle)$ and
$|\phi_\pm\rangle=\frac{1}{\sqrt{2}}(|01\rangle\pm|10\rangle)$ are Bell states. This 
result was verified by yet another numerical search procedure carried out recently \cite{Borras}. This
state exhibits genuine multi-partite entanglement according to
both negative partial transpose measure, as well as von Neumann
entropy measure. The von Neumann entropy between (1234$|$5) is
equal to one and between (123$|$45) is two. These are the maximum
possible entanglement values between the respective subsets, 
thus satisfying the general condition for teleportation as shown 
in \cite{Li}. It is also to be noted that $Tr(\rho_{i_{1}}^{2}) = \frac{1}{2}$ and
$Tr(\rho_{i_{1},i_{2}}^{2}) . . . = \frac{1}{4}$, where $i_1, i_2 . . .$ refer to the subsystems respectively,
thus satisfying the criteria for multiqubit
entanglement as shown in \cite{rigo}. The above state is also genuinely entangled according
to the recently proposed multiple entropy measures
(MEMS) \cite{MEMS}; it has MEMS of $S_1 = 1$ and $S_2 = 2$ respectively. This is more
than the entanglement exhibited by the GHZ, W and the cluster states.
Even
after tracing out one/two qubits from the state, entanglement
sustains in the resulting subsystem and thus, is highly `robust'.
Also, the state is maximally mixed, after we trace out any possible number of qubits, which is an indication of genuine
multi-particle entanglement for the five-qubit state
$|\psi_5\rangle$. Four-qubit states do not show such
characteristic behaviour and fail to attain maximal entropy
\cite{Higuchi}. Moreover, the above state assumes the same form 
for all 10 splits as (3+2). Thus , Alice can have any pair of 3 qubits 
in the above state to teleport to Bob.
 This is not possible with the four-qubit states known before.  Thus, the 
 five-qubit state can provide an edge over the four-qubit states for state
 transfer and coding. We shall now, devise a suitable method to 
 study the physical realization of this state and investigate its
 usefulness for quantum information tasks, namely teleportation
 state sharing and super-dense coding.  
\section{Physical realization}
Though the Brown state was initally obtained through an extensive numerical search
procedure, it can be physically realized as follows. We start with two photons in the Bell state given by
\begin{equation}
|\phi_{+}\rangle = \frac{1}{\sqrt{2}}(|01\rangle + |10\rangle).
\end{equation}  
We need to prepare another photon in the state $\frac{1}{\sqrt{2}}(|0\rangle+|1\rangle)$. One can combine both these states
and perform a $UC\ NOT$ operation on the last two qubits  and get a $W$ class of states as follows :
\begin{equation}
\frac{1}{2} (|01\rangle + |10\rangle) (|0\rangle+|1\rangle)
\stackrel{UC NOT(3,2)}{\rightarrow} \frac{1}{2}(|100\rangle+|010\rangle+|001\rangle+|111\rangle).
\end{equation}
We now take two photons in another Bell state, 
\begin{equation}
|\psi_+\rangle = \frac{1}{\sqrt{2}} (|00\rangle+|11\rangle).
\end{equation}   
The Brown state can be obtained by applying an unitary transformation $U_b$ to their combined state as follows :
\begin{equation}
|W\rangle \ |\psi_+\rangle \stackrel{U_b}{\rightarrow} |\psi_5\rangle 
\end{equation}
The unitary transform $U_b$ is given by a $32\times 32$ matrix, with unity in the places 
$
k_{1,1}, k_{2,2}, k_{5,6}, k_{6,5}, k_{9,9}, k_{10,10}, k_{13,14}, k_{14,13}, k_{17,18}, k_{18,17}, k_{19,20}, k_{20,19}, k_{21,21}, k_{23,23}, k_{24,24}, k_{25,26}, 
k_{26,25}, k_{27,28}, k_{28,27}\\, $$ k_{29,29}, k_{30,30}, k_{31,31}, k_{32,32}$
 and $-1$ in the following places 
$k_{3,3}, k_{4,4}, k_{7,8}, k_{8,7}, k_{11,11}, k_{12,12}, k_{15,16}, k_{16,15}$ and $0$ in the other terms. Here $k_{i,j}$ represents
the element in the $i^{th}$ row and $j^{th}$ column. This unitary operator can be further
decomposed into known gates in Quantum information. 
This procedure can lead to its possible experimental realization. 
\section{Teleportation of a single qubit state}
Let us first consider the situation in which Alice possesses
qubits $1$, $2$, $3$, $4$ and particle $5$ belongs to Bob. Alice
wants to teleport ($\alpha|0\rangle+\beta|1\rangle$) to Bob. So,
Alice prepares the combined state,

\begin{eqnarray}
(\alpha|0\rangle+\beta|1\rangle)|\psi_{5}\rangle=|\phi_{1}\rangle_{a_{1+}}
(\alpha|0\rangle+\beta|1\rangle)+|\phi_{2}\rangle_{a_{1-}}(\alpha|0\rangle-\beta|1\rangle)
+|\phi_{3}\rangle_{a_{2+}}(\beta|0\rangle+\alpha|1\rangle)+|\phi_{4}\rangle_{a_{2-}}(\beta|0\rangle-\alpha|1\rangle),
\end{eqnarray}

where, the $|\phi_{x}\rangle_{a_{i}\pm}$ are mutually orthogonal
states of the measurement basis. The states
$|\phi_{x}\rangle_{a_{i}\pm}$ are given as,
\begin{eqnarray}
|\phi_{x}\rangle_{a_{1}\pm}&=&(-|00011\rangle+|00100\rangle+|01001\rangle+|01110\rangle)\pm(|10010\rangle-
|10101\rangle+|11000\rangle \nonumber+|11111\rangle),\\
|\phi_{x}\rangle_{a_{2}\pm}&=&(-|10011\rangle+|10100\rangle+|11001\rangle+|11110\rangle)\pm(|00010\rangle-
|00101\rangle+|01000\rangle\nonumber+|01111\rangle).
\end{eqnarray}
Alice can now make a five-particle measurement using
$|\phi_{x}\rangle_{a_{i}\pm}$ and convey the outcome of her measurement to Bob via two classical bits. Bob can apply suitable unitary
operations given by $(1,\sigma_{1},i\sigma_{2},\sigma_{3})$ to
recover the original state $(\alpha|0\rangle+\beta|1\rangle)$.
This completes the teleportation protocol for the teleportation of
a single qubit state using the state $|\psi_5\rangle$. We now proceed to study the suitability of the Brown state for quantum state sharing (QSTS) of a single qubit state.

\section{QSTS of a single qubit state}
\subsection{Proposal I}
Let us consider the situation in which Alice possesses qubit 1,  Bob possess qubit 2, 3, 4 and Charlie 5. Alice has a unknown qubit ($\alpha|0\rangle+\beta|1\rangle$)
which she wants Bob and Charlie to share. Now, Alice combines the unknown 
qubit with the Brown state and performs a Bell measurement and convey her outcome to Charlie by two cbits. For instance, if Alice 
measures in the basis $|\psi_+\rangle$, then the Bob-Charlie system evolves into the entangled state:
\begin{equation}
\alpha(|01\rangle|\phi_{-}\rangle+|10\rangle|\psi_{-}\rangle)+\beta(|00\rangle|\phi_+\rangle +|11\rangle|\psi_+\rangle).
\end{equation}

Now Bob can perform a three partite measurement and convey his outcome to Charlie by two cbits. Having known the outcome of both
their measurements, Charlie can obtain the state by performing appropriate unitary transformations. The outcome of the
measurement performed by Bob and the state obtained by Charlie are shown in the table below:

 \begin{table}[h]
\caption{\label{tab1} The outcome of the measurement performed by Bob and the state obtained by Charlie}
\begin{tabular}{|c|c|}
\hline {\bf Outcome of the measurement } & {\bf State obtained }\\
$\frac{1}{2}(|010\rangle-|101\rangle +|001\rangle+|110\rangle)$&$\alpha|1\rangle+\beta|0\rangle$\\
$\frac{1}{2}(|100\rangle -|011\rangle +|000\rangle+|111\rangle)$&$\alpha|0\rangle+\beta|1\rangle$\\
$\frac{1}{2}(|010\rangle-|101\rangle -|001\rangle-|110\rangle)$&$\alpha|1\rangle-\beta|0\rangle$\\
$\frac{1}{2}(|100\rangle -|000\rangle-|111\rangle-|011\rangle)$&$\alpha|0\rangle-\beta|1\rangle$\\
\hline
\end{tabular}
\end{table}	
Here, Bob can also perform a single particle measurement followed by a two particle measurement
instead of a three particle measurement. However, this would consume an extra cbit of information.
\subsection{Proposal II}

In this scenario we let Alice possess qubit 1,2, Bob possess qubits 3,4 and Charlie 5. Alice combines the unknown qubit with
her particles and makes a three-partite measurement.  The outcome of the 
measurement performed by Alice and the entangled state obtained by Bob and Charlie are shown in the table below:

\begin{table}[h]
\caption{\label{tab1} The outcome of the measurement performed by Alice and the state obtained by Bob and Charlie}
\begin{tabular}{|c|c|}
\hline {\bf Outcome of the measurement } & {\bf State obtained }\\
$\frac{1}{4}(|000\rangle+|001\rangle +|110\rangle+|111\rangle)$&$\alpha(|1\rangle|\phi_{-}\rangle+|0\rangle|\psi_{-}\rangle)+\beta(|0\rangle|\phi_{+}\rangle+|1\rangle|\psi_{+}\rangle)$\\
$\frac{1}{4}(|000\rangle-|001\rangle -|110\rangle+|111\rangle)$&$\alpha(|1\rangle|\phi_{-}\rangle-|0\rangle|\psi_{-}\rangle)-\beta(|0\rangle|\phi_{+}\rangle-|1\rangle|\psi_{+}\rangle)$\\
$\frac{1}{4}(|000\rangle-|001\rangle +|110\rangle-|111\rangle)$&$\alpha(|1\rangle|\phi_{-}\rangle-|0\rangle|\psi_{-}\rangle)+\beta(|0\rangle|\phi_{+}\rangle-|1\rangle|\psi_{+}\rangle)$\\
$\frac{1}{4}(|000\rangle+|001\rangle -|110\rangle-|111\rangle)$&$\alpha(|1\rangle|\phi_{-}\rangle+|0\rangle|\psi_{-}\rangle)-\beta(|0\rangle|\phi_{+}\rangle+|1\rangle|\psi_{+}\rangle)$\\
$\frac{1}{4}(|100\rangle+|101\rangle +|010\rangle+|011\rangle)$&$\beta(|1\rangle|\phi_{-}\rangle+|0\rangle|\psi_{-}\rangle)+\alpha(|0\rangle|\phi_{+}\rangle+|1\rangle|\psi_{+}\rangle)$\\
$\frac{1}{4}(|100\rangle-|101\rangle -|010\rangle+|011\rangle)$&$\beta(|1\rangle|\phi_{-}\rangle-|0\rangle|\psi_{-}\rangle)-\alpha(|0\rangle|\phi_{+}\rangle-|1\rangle|\psi_{+}\rangle)$\\
$\frac{1}{4}(|100\rangle-|101\rangle +|010\rangle-|011\rangle)$&$\beta(|1\rangle|\phi_{-}\rangle-|0\rangle|\psi_{-}\rangle)+\alpha(|0\rangle|\phi_{+}\rangle-|1\rangle|\psi_{+}\rangle)$\\
$\frac{1}{4}(|100\rangle+|101\rangle -|010\rangle-|011\rangle)$&$\beta(|1\rangle|\phi_{-}\rangle+|0\rangle|\psi_{-}\rangle)-\alpha(|0\rangle|\phi_{+}\rangle+|1\rangle|\psi_{+}\rangle)$\\

\hline
\end{tabular}
\end{table}	
Alice can send the outcome of her measurement to Bob via three cbits of information.   
Now Bob and Charlie can meet and apply a joint unitary three particle transformations on their particles and convert it into the GHZ type of state as follows: 
\begin{equation}
\alpha(|1\rangle|\phi_{-}\rangle+|0\rangle|\psi_{-}\rangle)+\beta(|0\rangle|\phi_{+}\rangle+|1\rangle|\psi_{+}\rangle) \rightarrow \alpha|000\rangle+\beta|111\rangle.
\end{equation}
After performing the unitary transformation, Bob and Charlie can be spatially separated.
Bob can perform a Bell measurement on his particles and Charlie can obtain the state by applying an appropriate unitary operator.
This kind of statergy, might as well find applications other than state sharing in Quantum information. As in the previous case, even in
this scenario Alice can perform a Bell measurement followed by a single partite measurement instead of a three particle measurement. 
\section{Teleportation of an arbitrary two qubit state}
Alice has an arbitrary two qubit state,
\begin{equation}
|\psi\rangle=\alpha|00\rangle+\mu|10\rangle+\gamma|01\rangle+\beta|11\rangle,
\end{equation}

which she has to teleport to Bob. Here  $\alpha$, $\beta$, $\gamma$ and $\mu$ are any set of complex 
numbers satisfying $|\alpha|^2+|\beta|^2+|\gamma|^2+|\mu|^2 = 1$. Qubits $1$, $2$, $3$ and $4$, $5$
respectively, belong to Alice and Bob. Alice prepares the combined state,

\begin{eqnarray}
|\psi\rangle|\psi_{5}\rangle & = &
\frac{1}{4}[|\psi_{5}\rangle_1(\alpha|01\rangle+\gamma|00\rangle+\mu|11\rangle+\beta|10\rangle)+
|\psi_5\rangle_2(\alpha|01\rangle+\gamma|00\rangle-\mu|11\rangle-\beta|10\rangle)+\nonumber
\\ & &
|\psi_5\rangle_3(\alpha|01\rangle-\gamma|00\rangle+\mu|11\rangle-\beta|10\rangle)+
|\psi_5\rangle_4(\alpha|01\rangle-\gamma|00\rangle-\mu|11\rangle+\beta|10\rangle)+\nonumber
\\ & &
|\psi_5\rangle_5(\alpha|11\rangle+\gamma|10\rangle+\mu|01\rangle+\beta|00\rangle)+
|\psi_5\rangle_6(\alpha|11\rangle-\gamma|10\rangle+\mu|01\rangle-\beta|00\rangle)+\nonumber
\\ & &
|\psi_5\rangle_7(\alpha|11\rangle+\gamma|10\rangle-\mu|01\rangle-\beta|00\rangle)+
|\psi_5\rangle_8(\alpha|11\rangle-\gamma|10\rangle-\mu|01\rangle+\beta|00\rangle)+\nonumber
\\ & &
|\psi_5\rangle_9(\alpha|00\rangle+\gamma|01\rangle+\mu|10\rangle+\beta|11\rangle)+
|\psi_5\rangle_{10}(\alpha|00\rangle-\gamma|01\rangle+\mu|10\rangle-\beta|11\rangle)+\nonumber
\\ & &
|\psi_5\rangle_{11}(\alpha|00\rangle+\gamma|01\rangle-\mu|10\rangle-\beta|11\rangle)+
|\psi_5\rangle_{12}(\alpha|00\rangle-\gamma|01\rangle-\mu|10\rangle+\beta|11\rangle)+\nonumber
\\ & &
|\psi_5\rangle_{13}(\alpha|10\rangle+\gamma|11\rangle+\mu|00\rangle+\beta|01\rangle)+
|\psi_5\rangle_{14}(\alpha|10\rangle-\gamma|11\rangle+\mu|00\rangle-\beta|01\rangle)+\nonumber
\\ & &
|\psi_5\rangle_{15}(\alpha|10\rangle+\gamma|11\rangle-\mu|00\rangle-\beta|01\rangle)+
|\psi_5\rangle_{16}(\alpha|10\rangle-\gamma|11\rangle-\mu|00\rangle+\beta|01\rangle)].
\end{eqnarray}
Here, $|\psi_5\rangle_{i}$'s forming the mutual orthogonal basis
of measurement are given by :
\begin{eqnarray}
|\psi_{5}\rangle_{1}=\frac{1}{2}[|\phi_{-}\rangle|010\rangle+|\phi_{+}\rangle|111\rangle+|\psi_{-}\rangle|001\rangle+|\psi_{+}\rangle|
100\rangle];
|\psi_{5}\rangle_{2}=\frac{1}{2}[|\phi_{+}\rangle|010\rangle+|\phi_{-}\rangle|111\rangle+|\psi_{+}\rangle|001\rangle+|\psi_{-}\rangle|
100\rangle];\nonumber \\
|\psi_{5}\rangle_{3}=\frac{1}{2}[|\psi_{-}\rangle|001\rangle+|\psi_{+}\rangle|100\rangle-|\phi_{+}\rangle|010\rangle-|\phi_{-}\rangle|
111\rangle];
|\psi_{5}\rangle_{4}=\frac{1}{2}[|\psi_{-}\rangle|001\rangle+|\psi_{+}\rangle|100\rangle-|\phi_{-}\rangle|010\rangle-|\phi_{+}\rangle|
111\rangle];\nonumber \\
|\psi_{5}\rangle_{5}=\frac{1}{2}[|\psi_{+}\rangle|111\rangle-|\psi_{-}\rangle|010\rangle-|\phi_{-}\rangle|001\rangle+|\phi_{+}\rangle|
100\rangle];
|\psi_{5}\rangle_{6}=\frac{1}{2}[|\psi_{-}\rangle|111\rangle-|\psi_{+}\rangle|010\rangle+|\phi_{+}\rangle|001\rangle-|\phi_{-}\rangle|
100\rangle];\nonumber \\
|\psi_{5}\rangle_{7}=\frac{1}{2}[|\psi_{-}\rangle|111\rangle-|\psi_{+}\rangle|010\rangle-|\phi_{+}\rangle|001\rangle+|\phi_{-}\rangle|
100\rangle];
|\psi_{5}\rangle_{8}=\frac{1}{2}[|\psi_{+}\rangle|111\rangle-|\psi_{-}\rangle|010\rangle+|\phi_{-}\rangle|001\rangle-|\phi_{+}\rangle|
100\rangle];\nonumber \\
|\psi_{5}\rangle_{9}=\frac{1}{2}[|\psi_{+}\rangle|111\rangle+|\psi_{-}\rangle|010\rangle+|\phi_{-}\rangle|001\rangle+|\phi_{+}\rangle|
000\rangle];
|\psi_{5}\rangle_{10}=\frac{1}{2}[|\psi_{-}\rangle|111\rangle+|\psi_{+}\rangle|010\rangle-|\phi_{-}\rangle|001\rangle-|\psi_{+}\rangle|
100\rangle];\nonumber \\
|\psi_{5}\rangle_{11}=\frac{1}{2}[|\psi_{-}\rangle|111\rangle+|\psi_{+}\rangle|010\rangle+|\phi_{+}\rangle|001\rangle+|\phi_{-}\rangle|
100\rangle];
|\psi_{5}\rangle_{12}=\frac{1}{2}[|\psi_{+}\rangle|111\rangle+|\psi_{-}\rangle|010\rangle-|\phi_{+}\rangle|100\rangle-|\phi_{-}\rangle|
000\rangle];\nonumber \\
|\psi_{5}\rangle_{13}=\frac{1}{2}[|\psi_{+}\rangle|100\rangle-|\psi_{-}\rangle|001\rangle-|\phi_{-}\rangle|010\rangle+|\phi_{+}\rangle|
111\rangle];
|\psi_{5}\rangle_{14}=\frac{1}{2}[|\psi_{-}\rangle|100\rangle-|\psi_{+}\rangle|001\rangle+|\phi_{+}\rangle|010\rangle-|\phi_{-}\rangle|
111\rangle];\nonumber \\
|\psi_{5}\rangle_{15}=\frac{1}{2}[|\psi_{-}\rangle|100\rangle-|\psi_{+}\rangle|001\rangle-|\phi_{+}\rangle|010\rangle+|\phi_{-}\rangle|
111\rangle];
|\psi_{5}\rangle_{16}=\frac{1}{2}[|\psi_{+}\rangle|100\rangle-|\psi_{-}\rangle|001\rangle+|\phi_{-}\rangle|010\rangle-|\phi_{+}\rangle|
111\rangle].
\end{eqnarray}
Alice can make a five-particle measurement and then convey her
results to Bob. Bob then retrieves the original state
$|\psi\rangle_{b}$ by applying any one of the unitary transforms
shown in Table III to the respective states. As is evident,
each of the above states are obtained with equal probability. This
successfully completes the teleportation protocol of a two qubit
state using $|\psi_{5}\rangle$.
\begin{table}[h]
\caption{\label{tab1}Set of Unitary operators needed to obtain
$|\psi\rangle_b$}
\begin{tabular}{|c|c|}
\hline {\bf State} & {\bf Unitary}\\
& {\bf Operation}\\
\hline
$(\alpha|01\rangle+\gamma|00\rangle+\mu|11\rangle+\beta|10\rangle)$
& $I\otimes\sigma_{1}$\\
$(\alpha|01\rangle+\gamma|00\rangle-\mu|11\rangle-\beta|10\rangle)$
& $\sigma_{3}\otimes\sigma_{1}$\\
$(\alpha|01\rangle-\gamma|00\rangle+\mu|11\rangle-\beta|10\rangle)$&
$I\otimes i\sigma_{2}$\\
$(\alpha|01\rangle-\gamma|00\rangle-\mu|11\rangle+\beta|10\rangle)$&
$\sigma_{3}\otimes i\sigma_{2}$\\
$(\alpha|11\rangle+\gamma|10\rangle+\mu|01\rangle+\beta|00\rangle)$
& $\sigma_{1}\otimes\sigma_{1}$\\
$(\alpha|11\rangle-\gamma|10\rangle+\mu|01\rangle-\beta|00\rangle)$&
$\sigma_{1}\otimes i\sigma_{2}$\\
$(\alpha|11\rangle+\gamma|10\rangle-\mu|01\rangle-\beta|00\rangle)$&
$i\sigma_{2}\otimes\sigma_{1}$\\
$(\alpha|11\rangle-\gamma|10\rangle-\mu|01\rangle+\beta|00\rangle)$
& $i\sigma_{2}\otimes i\sigma_{2}$\\
$(\alpha|00\rangle+\gamma|01\rangle+\mu|10\rangle+\beta|11\rangle)$ & $I\otimes I$\\
$(\alpha|00\rangle-\gamma|01\rangle+\mu|10\rangle-\beta|11\rangle)$ & $I\otimes\sigma_{3}$\\
$(\alpha|00\rangle+\gamma|01\rangle-\mu|10\rangle-\beta|11\rangle)$
& $\sigma_{3}\otimes I$\\
$(\alpha|00\rangle-\gamma|01\rangle-\mu|10\rangle+\beta|11\rangle)$
& $\sigma_{3}\otimes\sigma_{3}$\\
$(\alpha|10\rangle+\gamma|11\rangle+\mu|00\rangle+\beta|01\rangle)$
& $\sigma_{1}\otimes I$\\
$(\alpha|10\rangle-\gamma|11\rangle+\mu|00\rangle-\beta|01\rangle)$
& $\sigma_{1}\otimes\sigma_{3}$\\
$(\alpha|10\rangle+\gamma|11\rangle-\mu|00\rangle-\beta|01\rangle)$
& $i\sigma_{2}\otimes I$\\
$
(\alpha|10\rangle-\gamma|11\rangle-\mu|00\rangle+\beta|01\rangle)$
&
$i\sigma_{2}\otimes\sigma_{3}$\\
\hline
\end{tabular}
\end{table}

\section{QSTS of an arbitrary two qubit state}
QSTS of an arbitrary two-particle
state was previously carried out using four Bell pairs among two controllers and then generalized to 
$N$ agents \cite{Deng}. We now demonstrate the utility of the Brown state for the QSTS of an arbitrary
two qubit state. It is evident that this protocol requires lesser number of
particles than the previously known protocol  and due to the properties
of the Brown state, the protocol is also more robust against decoherence. We propose 
one possible protocol for the QSTS of an arbitrary two qubit state.

We let Alice posses particles 1,2, Bob have particle 3, and 
Charlie has particles 4 and 5 in the Brown state respectively. Alice first,
combines the state $|\psi\rangle$ with the Brown state and makes a four - particle measurement.
The outcome of the measurement made by Alice and the entangled state obtained by Bob and Charlie are shown in the
table below :
\begin{table}[h]
\caption{\label{tab1} The outcome of the measurement performed by Alice and the state obtained by Bob
and Charlie(The coefficient $\frac{1}{4}$ is removed for convenience)}
\begin{tabular}{|c|c|}
\hline {\bf Outcome of the measurement } & {\bf State obtained}\\
\hline
$(|0000\rangle+|1001\rangle+|0110\rangle+|1111\rangle)$
& $\alpha|\Omega_1\rangle+\mu|\Omega_2\rangle+\gamma|\Omega_3\rangle+\beta|\Omega_4\rangle $\\
$(|0000\rangle-|1001\rangle+|0110\rangle-|1111\rangle)$
& $\alpha|\Omega_1\rangle-\mu|\Omega_2\rangle+\gamma|\Omega_3\rangle-\beta|\Omega_4\rangle $\\

$(|0000\rangle+|1001\rangle-|0110\rangle-|1111\rangle)$
& $\alpha|\Omega_1\rangle+\mu|\Omega_2\rangle-\gamma|\Omega_3\rangle-\beta|\Omega_4\rangle $\\

$(|0000\rangle-|1001\rangle-|0110\rangle+|1111\rangle)$
& $\alpha|\Omega_1\rangle-\mu|\Omega_2\rangle-\gamma|\Omega_3\rangle+\beta|\Omega_4\rangle $\\

$(|0010\rangle+|0101\rangle+|1000\rangle+|1101\rangle)$
& $\alpha|\Omega_3\rangle+\gamma|\Omega_4\rangle+\mu|\Omega_1\rangle+\beta|\Omega_2\rangle $\\
$(|0010\rangle-|0101\rangle+|1000\rangle-|1101\rangle)$
& $\alpha|\Omega_3\rangle-\gamma|\Omega_4\rangle+\mu|\Omega_1\rangle-\beta|\Omega_2\rangle $\\
$(|0010\rangle+|0101\rangle-|1000\rangle-|1101\rangle)$
& $\alpha|\Omega_3\rangle+\gamma|\Omega_4\rangle-\mu|\Omega_1\rangle-\beta|\Omega_2\rangle $\\
$(|0010\rangle-|0101\rangle-|1000\rangle+|1101\rangle)$
& $\alpha|\Omega_3\rangle-\gamma|\Omega_4\rangle-\mu|\Omega_1\rangle+\beta|\Omega_2\rangle $\\
$(|0001\rangle+|0100\rangle+|1011\rangle+|1110\rangle)$
& $\alpha|\Omega_2\rangle+\gamma|\Omega_1\rangle+\mu|\Omega_4\rangle+\beta|\Omega_3\rangle $\\
$(|0001\rangle-|0100\rangle+|1011\rangle-|1110\rangle)$
& $\alpha|\Omega_2\rangle-\gamma|\Omega_1\rangle+\mu|\Omega_4\rangle-\beta|\Omega_3\rangle $\\
$(|0001\rangle+|0100\rangle-|1011\rangle-|1110\rangle)$
& $\alpha|\Omega_2\rangle+\gamma|\Omega_1\rangle-\mu|\Omega_4\rangle-\beta|\Omega_3\rangle $\\
$(|0001\rangle-|0100\rangle-|1011\rangle+|1110\rangle)$
& $\alpha|\Omega_2\rangle-\gamma|\Omega_1\rangle-\mu|\Omega_4\rangle+\beta|\Omega_3\rangle $\\
$(|0011\rangle+|1010\rangle+|0101\rangle+|1100\rangle)$
& $\alpha|\Omega_4\rangle+\mu|\Omega_3\rangle+\gamma|\Omega_2\rangle+\beta|\Omega_1\rangle $\\
$(|0011\rangle-|1010\rangle+|0101\rangle-|1100\rangle)$
& $\alpha|\Omega_4\rangle-\mu|\Omega_3\rangle+\gamma|\Omega_2\rangle-\beta|\Omega_1\rangle $\\
$(|0011\rangle+|1010\rangle-|0101\rangle-|1100\rangle)$
& $\alpha|\Omega_4\rangle+\mu|\Omega_3\rangle-\gamma|\Omega_2\rangle-\beta|\Omega_1\rangle $\\
$(|0011\rangle-|1010\rangle-|0101\rangle+|1100\rangle)$
& $\alpha|\Omega_4\rangle-\mu|\Omega_3\rangle-\gamma|\Omega_2\rangle+\beta|\Omega_1\rangle $\\

\hline
\end{tabular}
\end{table}
\\where , 
\begin{eqnarray}
|\Omega_1\rangle = \frac{1}{{2}}(|101\rangle-|110\rangle),\\
|\Omega_2\rangle = \frac{1}{{2}}(|000\rangle-|011\rangle),\\
|\Omega_3\rangle = \frac{1}{{2}}(|001\rangle+|010\rangle),\\
|\Omega_4\rangle = \frac{1}{{2}}(|100\rangle+|111\rangle).\\
\end{eqnarray}
Neither Bob nor Charlie can reconstruct the original state $|\psi\rangle$ from the above states
by local operations.  Now Bob performs a measurement on his particle in the basis $\frac{1}{\sqrt{2}}(|0\rangle\pm|1\rangle)$. For 
instance if Bob gets the state ($\alpha|\Omega_1\rangle+\mu|\Omega_2\rangle+\gamma|\Omega_3\rangle+\beta|\Omega_4\rangle $)
then Charlie's particle evolves into any one of the following states:
$(\pm\alpha|\phi_-\rangle+\mu|\psi_-\rangle+\gamma|\phi_+\rangle\pm\beta|\psi_+\rangle)$. 

Alice sends the outcome of her measurement by four classical bits and Bob by one classical bit to Charlie.  Having known
the outcomes of both their measurements,  
Charlie can do appropriate Unitary transformations to get back the state $|\psi\rangle$. For instance,
if Charlie gets the above states, then he performs the  unitary transformations  ($\pm|11\rangle\langle\psi_+| + |10\rangle\langle\phi_+ |)
\pm (|00\rangle \langle\phi_{-}| + |01\rangle\langle\psi_{-}|$) on his two particles to get back $|\psi\rangle$. Another possible scenario is that, Alice can send the result of her measurement to Bob 
by four classical bits of information. Then Bob and Charlie can co-operate and apply joint unitary transformations on their particles
and convert their state into ($\alpha|\Omega_1\rangle+\mu|\Omega_2\rangle+\gamma|\Omega_3\rangle+\beta|\Omega_4\rangle $) and then
be spatially separated. From, now on
the protocol follows the previous scenario.\\ Suppose, we let Alice have particle 1, Bob have particles 2,3 and 
Charlie have particles 4 and 5 in the Brown state. Alice can 
combine the state $|\psi\rangle$ with the Brown state and make a three - particle measurement. Thus, the Bob-Charlie system is left with four qubits. But, it is not possible to obtain the state $|\psi\rangle$ from their combined state,
in a straightforward manner. The Brown state could also be used for QSTS of a three partite GHZ type state given by : 
$|GHZ\rangle = \alpha|000\rangle + \beta|111\rangle$. \\These results could be directly generalized to $N$ agents using the following
state :
\begin{equation}
|\psi_{n}\rangle=\frac{1}{2}(|\eta_1\rangle_{n}|001\rangle|\phi_{-}\rangle+|\eta_2\rangle_{n}|010\rangle|\psi_{-}\rangle
+|\eta_3\rangle_{n}|100|\phi_{+}\rangle+|\eta_4\rangle_{n}|111\rangle|\psi_{+}\rangle),
\end{equation}
where the $|\eta_i\rangle$'s  form the computational basis
of the nth order. For example if n=2, $|\eta_i\rangle$ equals any combination from the set, $(|00\rangle, |11\rangle, |10\rangle, |01\rangle)$
This we call as the "generalized Brown state". We let Alice  have the first two particles, Charlie have last two particles and the other agents
in the network have the remaining particles. 

\section{Superdense coding}
We now proceed to show the utility of $|\psi_{5}\rangle$ for
superdense coding. Entanglement is quite handy in communicating
information efficiently, in a quantum channel. Suppose Alice and
Bob share an entangled state, namely $|\psi\rangle_{AB}$. Then Alice can
convert her state into different orthogonal states by applying
suitable unitary transforms on her particle \cite{Wiesner}. Bob
then does appropriate Bell measurements on his qubits to retrieve
the encoded information. It is known that two classical bits per
qubit can be exchanged by sending information through a Bell
state. In this section, we shall discuss the suitability of
$|\psi_5\rangle$, as a resource for superdense coding. Let us
assume that Alice has first three qubits, and Bob has last two
qubits. Alice can apply the set of unitary transforms on her
particle and generate $64$ states out of which $32$ are mutually
orthogonal as shown below:
\begin{equation}
U^{3}_{x}\otimes I\otimes I\rightarrow|\psi_{5}\rangle_{x_{i}}.
\end{equation}
Bob can then perform a five-partite measurement in the basis of
$|\psi_{5}\rangle_{x_{i}}$ and distinguish these states. The
appropriate unitary transforms applied and the respective states
obtained by Alice are shown in the Table \ref{tab2}.
\begin{table}[h]
\caption{\label{tab2} States $|\psi_{5}\rangle_{x_{i}}$ obtained
by Alice after performing unitary operations $U^{3}_{x}$}
\begin{tabular}{|c|c|}
\hline {\bf Unitary Operation} & {\bf State} \\
\hline
$I\otimes I\otimes I$ &
$\frac{1}{2}(|001\rangle|\phi_{-}\rangle+|010\rangle|\psi_{-}\rangle+|100\rangle|\phi_{+}\rangle+|111\rangle
|\psi_{+}\rangle)$\\
$I\otimes\sigma_{3}\otimes I$ &
$\frac{1}{2}(|001\rangle|\phi_{-}\rangle-|010\rangle|\psi_{-}\rangle+|100\rangle|\phi_{+}\rangle-|111\rangle
|\psi_{+}\rangle)$\\
$\sigma_{3}\otimes I\otimes I$ &
$\frac{1}{2}(|001\rangle|\phi_{-}\rangle+|010\rangle|\psi_{-}\rangle-|100\rangle|\phi_{+}\rangle-
|111\rangle|\psi_{+}\rangle)$\\
$\sigma_{3}\otimes \sigma_{3}\otimes I$ &
$\frac{1}{2}(|001\rangle|\phi_{-}\rangle-|010\rangle|\psi_{-}\rangle-|100\rangle|\phi_{+}\rangle
+|111\rangle|\psi_{+}\rangle)$\\
$\sigma_{1}\otimes \sigma_{1}\otimes I$ & $\frac{1}{2}(|111\rangle|\phi_{-}\rangle+|100\rangle|\psi_{-}\rangle+|010\rangle|\phi_{+}\rangle
+|001\rangle|\psi_{+}\rangle)$\\
$\sigma_{1}\otimes i\sigma_{2}\otimes I$& $\frac{1}{2}(|111\rangle|\phi_{-}\rangle-|100\rangle|\psi_{-}\rangle+|010\rangle|\phi_{+}\rangle
-|001\rangle|\psi_{+}\rangle)$\\
$i\sigma_{2}\otimes \sigma_{1}\otimes I$ &
$\frac{1}{2}(|111\rangle|\phi_{-}\rangle+|100\rangle|\psi_{-}\rangle-|010\rangle|\phi_{+}\rangle
-|001\rangle|\psi_{+}\rangle)$\\
$i\sigma_{2}\otimes i\sigma_{2}\otimes I$ &
$\frac{1}{2}(|111\rangle|\phi_{-}\rangle-|100\rangle|\psi_{-}\rangle-|010\rangle|\phi_{+}\rangle
+|001\rangle|\psi_{+}\rangle)$\\
$I\otimes\sigma_{1}\otimes I$ & $\frac{1}{2}(|011\rangle|\phi_{-}\rangle+|000\rangle|\psi_{-}\rangle+|110\rangle|\phi_{+}\rangle
+|101\rangle|\psi_{+}\rangle)$\\
$I\otimes i \sigma_{2}\otimes I$ & $\frac{1}{2}(|011\rangle|\phi_{-}\rangle-|000\rangle|\psi_{-}\rangle+|110\rangle|\phi_{+}\rangle
-|101\rangle|\psi_{+}\rangle)$\\
$\sigma_{3}\otimes\sigma_{1}\otimes I$ & $\frac{1}{2}(|011\rangle|\phi_{-}\rangle+|000\rangle|\psi_{-}\rangle-|110\rangle|\phi_{+}\rangle-
|101\rangle|\psi_{+}\rangle)$\\
$\sigma_{3}\otimes i\sigma_{2}\otimes I$ & $\frac{1}{2}(|000\rangle|\psi_{-}\rangle-|011\rangle|\phi_{-}\rangle-|110\rangle|\phi_{+}\rangle
+|101\rangle|\psi_{+}\rangle)$\\
$\sigma_{1}\otimes I\otimes I$ & $\frac{1}{2}(|101\rangle|\phi_{-}\rangle+|110\rangle|\psi_{-}\rangle+|000\rangle|\phi_{+}\rangle
+|011\rangle|\psi_{+}\rangle) $\\
$\sigma_{1} \otimes \sigma_{3}\otimes I$ & $\frac{1}{2}(|100\rangle|\phi_{-}\rangle-|110\rangle|\psi_{-}\rangle+|000\rangle|\phi_{+}
\rangle-|011\rangle|\psi_{+}\rangle)$\\
$i\sigma_{2}\otimes I\otimes I$ & $\frac{1}{2}(|000\rangle|\phi_{+}\rangle-|101\rangle|\phi_{-}\rangle-|110\rangle|\psi_{-}\rangle
+|011\rangle|\psi_{+}\rangle)$\\
$i\sigma_{2}\otimes\sigma_{3}\otimes I$ & $ \frac{1}{2}(|101\rangle|\phi_{-}\rangle+|110\rangle|\psi_{-}\rangle-|000\rangle|\phi_{+}\rangle
-|011\rangle|\psi_{+}\rangle)$\\
$I\otimes I\otimes \sigma_{1}$ & $\frac{1}{2}(|000\rangle|\phi_{-}\rangle+|011\rangle|\psi_{-}\rangle+|101\rangle|\phi_{+}\rangle
+|110\rangle|\psi_{+}\rangle) $\\
$I\otimes\sigma_{3}\otimes \sigma_{1}$ & $\frac{1}{2}(|000\rangle|\phi_{-}\rangle-|011\rangle|\psi_{-}\rangle+|101\rangle|\phi_{+}\rangle
-|110\rangle|\psi_{+}\rangle)$\\
$\sigma_{3}\otimes \sigma_{1}\otimes \sigma_{1}$ & $\frac{1}{2}(|000\rangle|\phi_{-}\rangle+|011\rangle|\psi_{-}\rangle-|101\rangle
|\phi_{+}\rangle-|110\rangle|\psi_{+}\rangle)$\\
$\sigma_{3}\otimes \sigma_{3}\otimes \sigma_{1}$ & $\frac{1}{2}(|000\rangle|\phi_{-}\rangle-|011\rangle|\psi_{-}\rangle-|101\rangle
|\phi_{+}\rangle+|110\rangle|\psi_{+}\rangle)$\\
$\sigma_{1}\otimes \sigma_{1}\otimes \sigma_{1}$ & $\frac{1}{2}(|110\rangle|\phi_{-}\rangle+|101\rangle|\psi_{-}\rangle+|011\rangle
|\phi_{+}\rangle+|000\rangle|\psi_{+}\rangle)$\\
$\sigma_{1}\otimes i\sigma_{2}\otimes \sigma_{1}$ & $\frac{1}{2}(|110\rangle|\phi_{-}\rangle-|100\rangle|\psi_{-}\rangle+|011\rangle
|\phi_{+}\rangle-|000\rangle|\psi_{+}\rangle)$\\
$\i\sigma_{2}\otimes \sigma_{1}\otimes \sigma_{1}$ & $\frac{1}{2}(|110\rangle|\phi_{-}\rangle+|101\rangle|\psi_{-}\rangle-|011\rangle
|\phi_{+}\rangle-|000\rangle|\psi_{+}\rangle)$\\
$i\sigma_{2}\otimes i\sigma_{2}\otimes \sigma_{1}$ & $\frac{1}{2}(|110\rangle|\phi_{-}\rangle-|101\rangle|\psi_{-}\rangle-|011\rangle
|\phi_{+}\rangle+|000\rangle|\psi_{+}\rangle)$\\
$I\otimes\sigma_{1}\otimes \sigma_{1}$ & $\frac{1}{2}(|010\rangle|\phi_{-}\rangle+|001\rangle|\psi_{-}\rangle+|111\rangle|\phi_{+}\rangle
+|100\rangle|\psi_{+}\rangle)$\\
$I\otimes i \sigma_{2}\otimes \sigma_{1}$ & $ \frac{1}{2}(|010\rangle|\phi_{-}\rangle-|001\rangle|\psi_{-}\rangle+|111\rangle
|\phi_{+}\rangle-|100\rangle|\psi_{+}\rangle)$\\
$\sigma_{3}\otimes\sigma_{1}\otimes \sigma_{1}$ & $\frac{1}{2}(|001\rangle|\psi_{-}\rangle-|010\rangle|\phi_{-}\rangle+|111\rangle
|\phi_{+}\rangle-|100\rangle|\psi_{+}\rangle)$ \\
$\sigma_{3}\otimes i\sigma_{2}\otimes \sigma_{1}$ & $\frac{1}{2}(|001\rangle|\psi_{-}\rangle-|010\rangle|\phi_{-}\rangle
-|111\rangle|\phi_{+}\rangle+|100\rangle|\psi_{+}\rangle)$\\
$\sigma_{1}\otimes I \otimes \sigma_{1}$ & $\frac{1}{2}(|100\rangle|\phi_{-}\rangle+|111\rangle|\psi_{-}\rangle
+|001\rangle|\phi_{+}\rangle+|010\rangle|\psi_{+}\rangle)$\\
$\sigma_{1} \otimes \sigma_{3}\otimes \sigma_{1}$ & $\frac{1}{2}(|100\rangle|\phi_{-}\rangle-|111\rangle|\psi_{-}\rangle
+|001\rangle|\phi_{+}\rangle-|010\rangle|\psi_{+}\rangle)$\\
$i\sigma_{2}\otimes I\otimes \sigma_{1}$ & $\frac{1}{2}(|001\rangle|\phi_{+}\rangle-|100\rangle|\phi_{-}\rangle-|111\rangle|\psi_{-}\rangle
+|010\rangle|\psi_{+}\rangle)$\\
$i\sigma_{2}\otimes\sigma_{3}\otimes \sigma_{1}$ & $\frac{1}{2}(|100\rangle|\phi_{-}\rangle+|111\rangle|\psi_{-}\rangle-
|001\rangle|\phi_{+}\rangle-|010\rangle|\psi_{+}\rangle)$\\
\hline
\end{tabular}
\end{table}
\\The capacity of superdense coding is defined as \cite{Bruss},
\begin{equation}
X(\rho^{AB})=\mathrm{log_{2}}d_{A}+S(\rho^{B})-S(\rho^{AB}),
\end{equation}
where $d_{A}$ is the dimension of Alice's system, $S(\rho)$ is
von-Neumann entropy. For the state $|\psi_{5}\rangle$,
$X(\rho^{AB})=3+2-0=5$. The Holevo bound of a multipartite quantum
state gives the maximum amount of classical information that can
be encoded \cite{Bruss}. It is equal to five, for the five-qubit
state ($\mathrm{log_{2}}N$). Thus, the super dense coding reaches the
"Holevo bound" allowing five classical bits to be transmitted
through three quantum bits consuming only two ebits of entanglement.
The Brown state could be used to send two classical bits
by sending a qubit consuming one ebit of entanglement. Hence,
the Brown state could be used instead of the Bell state 
considered in \cite{Wiesner}. 
One can also send four classical bits by sending two qubits consuming
two ebits of entanglement. 
Thus the Brown state could also be used instead of four partite cluster state considered in \cite{4pati}.
It could be shown that, using
the generalized Brown state, it is possible to send, $(2N-1)$ qubits
by sending $N$  classical bits, if N is odd, or else send $2N$ qubits
by sending $N$ classical bits if $N$ is even. Thus satisfying the definition
of TMES for superdense coding \cite{Pankaj}.\\
It is worth mentioning that, all the calculations in the paper with regard to teleportation,
state sharing and superdense coding could be carried out using the following state:
\begin{equation}
|\psi_{5}\rangle=\frac{1}{2}(|\Omega_1\rangle|\phi_{-}\rangle+|\Omega_2\rangle|\psi_{-}\rangle
+|\Omega_3\rangle|\phi_{+}\rangle+|\Omega_4\rangle|\psi_{+}\rangle).
\end{equation}
where $|\Omega_i\rangle$'s form a tri-partite orthogonal basis. However, the Brown state
makes it possible for Alice to have any three particles because it has the
same form for all $(3+2)$ splits \cite{sudbery}. All the applications 
considered in this paper could also be carried out using a state of the type :
\begin{equation}
|\psi_{5}\rangle = A_1|001\rangle|\phi_{-}\rangle+A_2|010\rangle|\psi_{-}\rangle
+A_3|100\rangle|\phi_{+}\rangle+A_4|111\rangle|\psi_{+}\rangle,
\end{equation}
where $A_i$ is an integer, if the following relations are satisfied :
\begin{eqnarray}
-\sum_{n=1}^{4} A_i^2(1+\mathrm{log_{2}} A_i^2) = 1, \\ -(A_3^2+A_4^2) \mathrm{log_{2}} (A_3^2+A_4^2) = \frac{1}{2}.
\end{eqnarray}
This is a necessary but not sufficient condition.  
\section{Conclusion}
We have shown that the new five partite state obtained by Brown
{\it et al.}, \cite{Brown} has many useful applications in quantum
information. We show that this state can be physicaly realized by a pair of 
Bell states and a single qubit state. We use this state for perfect teleportation and 
Quantum state sharing 
of an arbitrary one qubit and two-qubit states under different scenarios. This state is also a
very useful resource for superdense coding. The superdense
capacity for the state reaches Holevo bound of five classical
bits. The state under consideration helps one to carry out
teleportation and superdense coding maximally. In future, we wish
to generalize these protocols for other states having odd number
of qubits and qudits. The decoherence property of this state also
needs careful investigation in case of any practical applications. The
comparison between the cost function and decoherence properties
of different classes of states in Eq(20) and the Brown state can also be an interesting
future work. A series of papers will follow.
\begin{acknowledgments}
This work was supported by the NIUS program undertaken by the Homi Bhabha Centre
for Science Education (HBCSE - TIFR), Mumbai, India. The authors
acknowledge Prof. V. Singh of HBCSE for discussion and continuous
encouragement and Sidharth K for discussions. We thank Prof. A. Sudbery for his careful reading of the manuscript, comments and also for  
bringing the typographical error made by Brown \it{et al.}, \normalfont to our notice. 
\end{acknowledgments}

\end{document}